\documentclass[sigconf]{acmart}

\usepackage{stfloats} 

\AtBeginDocument{%
  }

\copyrightyear{2025}
\acmYear{2025}
\setcopyright{rightsretained}
\acmConference[RecSys '25]{Proceedings of the Nineteenth ACM Conference on Recommender Systems}{September 22--26, 2025}{Prague, Czech Republic}
\acmBooktitle{Proceedings of the Nineteenth ACM Conference on Recommender Systems (RecSys '25), September 22--26, 2025, Prague, Czech Republic}
\acmDOI{10.1145/3705328.3748139}
\acmISBN{979-8-4007-1364-4/2025/09}

\begin{document}

\title{Semantic IDs for Music Recommendation}

\author{M. Jeffrey Mei}
\email{jeffrey.mei@siriusxm.com}
\orcid{0000-0002-1083-598X}
\affiliation{%
  \institution{SiriusXM Radio Inc.}
  \city{New York}
  \state{New York}
  \country{USA}
}

\author{Florian Henkel}
\email{florianh@spotify.com}
\orcid{0009-0005-5212-3357}
\affiliation{%
  \institution{Spotify}
  \city{New York}
  \state{New York}
  \country{USA}
}

\author{Samuel E. Sandberg}
\orcid{0009-0006-7665-4766}
\author{Oliver Bembom}
\orcid{0009-0002-2617-5776}
\author{Andreas F. Ehmann}
\email{first.last@siriusxm.com}
\orcid{0009-0003-9589-0666}
\affiliation{%
  \institution{SiriusXM Radio Inc.}
  \city{New York}
  \state{New York}
  \country{USA}
}

\renewcommand{\shortauthors}{Mei et al.}

\begin{abstract}
Training recommender systems for next-item recommendation often requires unique embeddings to be learned for each item, which may take up most of the trainable parameters for a model.  Shared embeddings, such as using content information, can reduce the number of distinct embeddings to be stored in memory. 
This allows for a more lightweight model; correspondingly, model complexity can be increased due to having fewer embeddings to store in memory.
We show the benefit of using shared content-based features (`semantic IDs') in improving recommendation accuracy and diversity, while reducing model size, for two music recommendation datasets, including an online A/B test on a music streaming service.
\end{abstract}




\maketitle

\section{Introduction}
Music streaming companies often have vast catalogs with tens of millions of distinct tracks, of which only a small portion is actually played to each user. Recommendation models are necessary to identify and personalize this subset of songs for each user. 

These models need to span the catalog, and so may become impracticably large for real-time use. If a distinct embedding of hidden dimension $h$ is learned for each covered item within a catalog of size $N$, 
then $N \times h$ parameters are needed for item representation alone, before even including the model weights. 
This poses issues for training, as a model with too many items may not fit into memory and $h$ may be constrained by memory limitations. Such large models may also be cost-prohibitive for real-time inference.
%, and such large models may have prohibitively slow inference times for real-time inference.

There has been some research into reducing the dimensionality of learned item embeddings. 
Hashing is one such way of assigning items to a shared space. Hashing has some risk of item collision and poor generalization (for out-of-vocabulary items) due to the pseudo-randomness of the hashing process, as similar songs (for example, by the same artist) may not have similar hash IDs \cite{weinberger2009, shi2020b}. 
There is some evidence that this may actually improve generalization for \textit{covered} items by forcing model regularization \cite{lan2020, petrov2024}.

Another way is to factorize the embeddings, which can be done before or after training \cite{lan2020, kang2020}. 
One approach is to factorize the user-item matrix before training \cite{petrov2024}.
However, in this case, the model cannot generalize to new items that lack feedback. Factorizing after training means that the model complexity is still constrained at training time and may therefore yield an underpowered model.
%Factorizing (or otherwise reducing the dimensionality of) the item embeddings \textit{before} training would allow for higher item coverage for a given number of model parameters. 

If content features are available, they can be used to represent items within a shared semantic space \cite{singh2024} or via a two-tower method \cite{yi2019}. 
These `semantic IDs' have been used in industry-scale recommender models \cite{singh2024}. 
Similar to \cite{singh2024}, we experiment with content-based semantic IDs, but in a music recommender system. 
We compare with using randomly-generated IDs and find that trained semantic IDs outperform random IDs when used to replace song embeddings, but not  when additional song metadata are used. 

We evaluate this on two datasets, including one open-source dataset from Spotify, and conduct an online test on Pandora users.
For our key contributions, we show that:

% Relatedly, Petrov's work on factorized item embeddings using user feedback has the same intended effect of parameter reduction. 
% The authors also find that using random indices (which would be equivalent to using random indices for semantic IDs) also achieve  good performance, speculating this is due to improved regularization by forcing popular and rare items to share embeddings.

\begin{itemize}
    \item semantic IDs can reduce the number of trainable parameters without loss in accuracy; equivalently, these parameters can be used to make more complex models with better accuracy,
    % \item semantic IDs can be used to cover cold-start songs with reasonable accuracy, although once feedback exists for these songs, it is better to train bespoke embeddings,
    \item although accuracy generally increases with more trainable parameters, it may be better to increase the semantic codebook size instead of the number of  hidden dimensions,
    \item semantic IDs have the most increase in recommendation accuracy and diversity for low-feedback users
\end{itemize}
% \section{Data} \label{sect:data}
\enlargethispage*{16pt}

\section{Model and Datasets}
Two datasets with user feedback for online radio are used:  Spotify's open-source Sequential Skip Prediction dataset  \cite{brost2019} and a proprietary dataset from Pandora. 
Both are filtered for `radio stations' only and include positive and negative feedback. 
The Spotify dataset uses implicit play (positive) and skip (negative) feedback, while the Pandora dataset uses explicit thumb-up (positive) and thumb-down (negative) feedback.
Summary statistics are given in Table \ref{table:data}.
Pandora radio stations are seeded using a track, artist or genre, which is used to identify relevant songs.
\begin{table} [t]
\caption{\label{table:data}Summary statistics for the processed training sets. 
% $+$' denotes explicit positive feedback, `$-$' denotes explicit negative feedback and `$/$' denotes song skips. 
Statistics are not necessarily representative of the user base.}
\begin{tabular}{c|ccc}
\toprule
                         &     Spotify      & Pandora     \\ \midrule
Granularity            & Session      & User     \\
Feedback types        & play, skip     & up, down \\
Max. seq. length          & 20           & 400     \\
Max. lookback       &  Same day   & 1 year    \\
No. of sequences     & $8 \times 10^6$   & $10^7 $  \\
No. of tracks      & 3 $\times 10^6$ & $10^6$ \\ \bottomrule
\end{tabular}
\end{table}
% \vspace*{-\baselineskip} 

The baseline model for the Pandora dataset is a transformer model based off SASRec \cite{kang2018} with small modifications \cite{mei2023, mei2024}. We decompose the track embedding into artist and genre components, i.e. we define the `track' embedding as the sum of `song', `artist' and `genre' embeddings. 
Because the Spotify dataset does not include artist/genre information, the tracks are used directly by themselves, i.e. the `track' embedding is simply the `song' embedding. 
For a fairer comparison to the Spotify dataset, we train Pandora baselines both \textbf{with} and \textbf{without} song decomposition, as the artist/genre tags can provide considerable shared, content-based information.

The baseline and semantic models are trained identically, except for the treatment of the song embedding. 
For semantic models, the song embedding is decomposed into a $n$-dimensional semantic ID with codebook size $k$, while the baseline is simply a song index. 
This means the baseline is equivalent to a `one-dimensional' semantic ID with a codebook size equal to the number of covered songs.

\enlargethispage*{16pt}

\subsection{Semantic IDs}

\ifdim\columnwidth<0.7\textwidth

\begin{figure*}[b!]
 \includegraphics[width=0.86\textwidth]{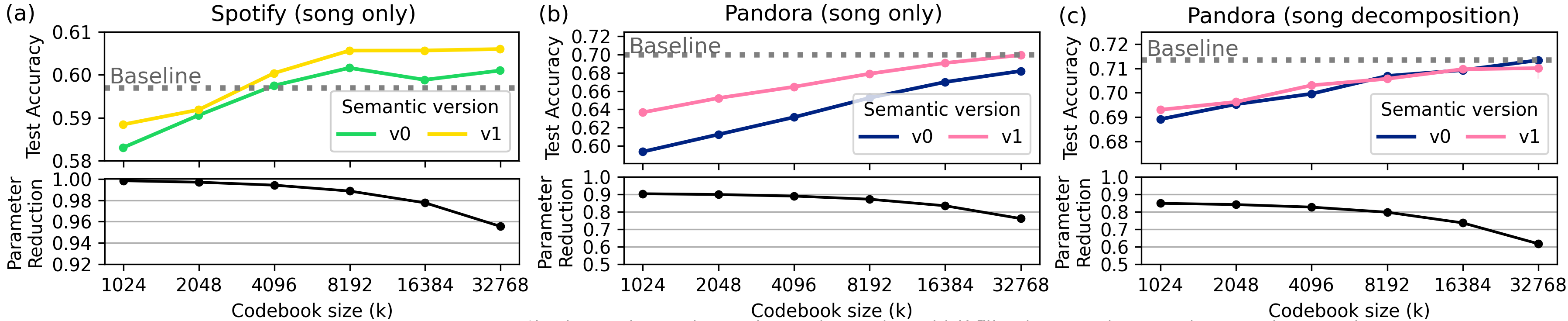}
\caption{Summarized results for the Spotify (a) and Pandora (b, c) datasets. In each subplot, the baseline (non-semantic) model accuracy is shown with a dotted gray line. As the codebook size $k$ increases, the test accuracy increases and the number of trainable parameters also increases (i.e. less parameter reduction). The trained (`v1') semantic IDs outperform the randomly-assigned (`v0') semantic IDs for the song-only case (a, b), but show little benefit when using additional song metadata (c).}
  \label{fig:results_merged}
\Description{A figure showing the accuracy of semantic ID models generally increases with increasing codebook size; for Spotify this plateaus after around k=8192 (panel a). The song-only models show much higher accuracy for v1 vs v0 (panels a and b), whereas there is almost no difference between them when using song decomposition (panel c). In all cases, the number of parameters is reduced when using semantic IDs; this reduction can be of order 98\% for Spotify (codebook size 16384) and order 75\% for Pandora (codebook size 16384)}
\end{figure*}
\else

\fi

The main idea behind semantic IDs is to represent recommendable items as a $n$-tuple of codewords, which come from separate codebooks each of size $k$. We use $n=4$ and vary $k$; for example, with $k=64$, this can encode $64^4$ ($\sim$ 16M) unique items. 

One simpler alternative is to randomly assign IDs for each item (hereafter `v0'). This is somewhat similar to the random hashing trick. 
However, this completely ignores the underlying information of the items and does not impose any structure. 
On the contrary, the goal of semantic IDs is to use item information, such as content features/embeddings, such that similar items end up sharing semantic IDs. 
For example, songs by the same artists or a similar genre may share the same first (few) codewords.

Similar to \cite{rajput2023}, we use a residual variational auto-encoder (RQ-VAE) to create the semantic IDs \cite{zeghidour2022}. We append an additional dimension for tie-breaking (i.e. $n+1$ dimensions in total). This final ID is not learned, but rather incremented whenever two songs have the same  first four (learned) semantic IDs. 
Such collisions may occur if some items have almost-indistinguishable content features. 
These \textit{trained} semantic IDs (hereafter `v1'), are contrasted to the randomly-assigned `v0' IDs, which may nevertheless provide insights for other datasets that lack content embeddings. 

The semantic IDs for the Spotify dataset are trained using an 8-dimensional `audio vector' and additional content attributes provided in the Spotify Sequential Skip Prediction dataset \cite{brost2019}. This includes track attributes such as `speechiness', `danceability', and `energy'. We exclude `popularity', which is not a content feature \cite{brost2019}.  
For Pandora, a proprietary audio embedding and embeddings derived from song metadata (e.g. genre, release year) are used \cite{mccallum2022, oramas2018}.

\section{Evaluation}
We evaluate the accuracy of ranking correctly the songs in a (positive, negative) feedback pair from the same user (Pandora) or same session (Spotify).
The test set is time-separated from the training/validation and is evaluated on feedback from the subsequent two weeks after the training period \cite{hidasi2023, wilm2023}.
This means that only users/sessions with at least one positive and one negative feedback from the test period are included. 
Additionally, the model input (excluding the positive/negative pair) must be non-empty. %, meaning that user/sessions must have at least one additional feedback in addition to the (positive, negative) pair.
% For Pandora, the positive and negative feedback is explicit up and down thumbs; for Spotify it is implicit play and skip (Sect. \ref{sect:data}).
This test accuracy, averaged over all users/sessions, is equivalent to a stratified area under the receiver operating curve (stratified AUC).  
For \textbf{each} of the three cases (Spotify, Pandora song-only, Pandora song-decomposition), we train a baseline with no semantic IDs.

\section{Results / Discussion}

\enlargethispage*{12pt}

\subsection{Offline evaluation}
For the Spotify dataset (Fig. \ref{fig:results_merged}a), we find that a codebook size of 4096 is able to match the baseline, and a codebook size of 8192 is able to beat the baseline, when using an otherwise-identical model architecture, for both `v0' (random) and `v1' (trained) semantic IDs.
For the Pandora dataset, we find that the song-only model (Fig. \ref{fig:results_merged}b), which is analogous to the Spotify model, can match the baseline using `v1' semantic IDs with a codebook size of 32768, with an otherwise-identical model (Fig. \ref{fig:results_merged}b). 

 \ifdim\columnwidth<0.7\textwidth
\else
\begin{figure*}[h]
 \includegraphics[width=0.86\textwidth]{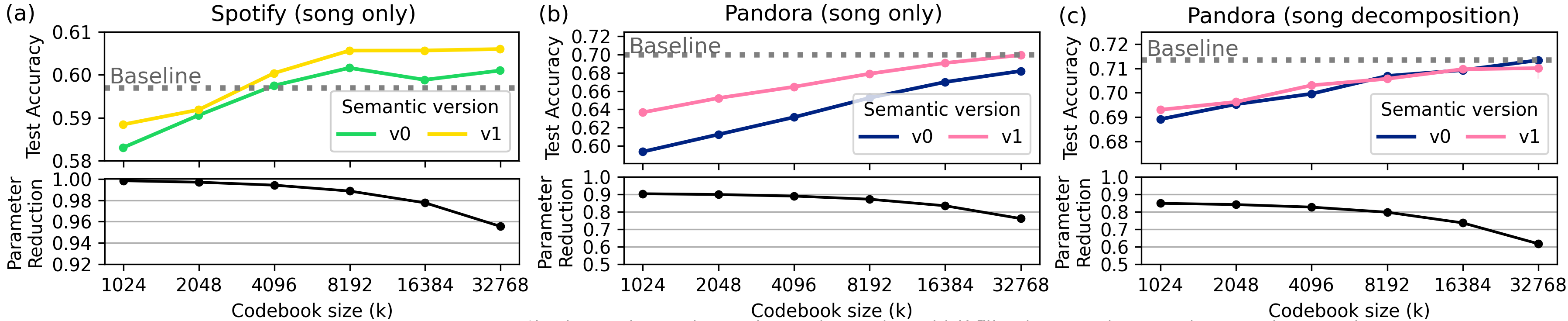}
% \begin{figure}[h]
%  \includegraphics[width=0.5\textwidth]{figs/results_merged_half.png}
\caption{Summarized results for the Spotify (a) and Pandora (b, c) datasets. In each subplot, the baseline (non-semantic) model accuracy is shown with a dotted gray line. As the codebook size $k$ increases, the test accuracy increases and the number of trainable parameters also increases (i.e. less parameter reduction). The trained (`v1') semantic IDs outperform the randomly-assigned (`v0') semantic IDs for the song-only case (a, b), but show little benefit when using additional song metadata (c).}
  \label{fig:results_merged}
  \Description{A figure showing the accuracy of semantic ID models generally increases with increasing codebook size; for Spotify this plateaus after around k=8192 (panel a). The song-only models show much higher accuracy for v1 vs v0 (panels a and b), whereas there is almost no difference between them when using song decomposition (panel c). In all cases, the number of parameters is reduced when using semantic IDs; this reduction can be of order 98\% for Spotify (codebook size 16384) and order 75\% for Pandora (codebook size 16384)}
\end{figure*}
\fi

In both cases, this greatly reduces the number of trainable parameters (by $\sim$75\% for Pandora and 99\% for Spotify). 
This allows for additional trainable parameters to be assigned to increasing model complexity, e.g. with more hidden dimensions (see Sect. \ref{sec:varying}).

The song-only results also show that the `v1' semantic IDs have higher accuracy than the `v0' semantic IDs, for both the song-only Pandora and Spotify datasets (Fig. \ref{fig:results_merged}a, b). This gap in performance decreases with higher codebook sizes; indeed, the baseline model is equivalent to a (one-dimensional) codebook of size $N_{\text{songs}}$, and if there are no collisions, then at this limit `v1' and `v0' are  equivalent. 

\subsubsection{Song decomposition}
When using the song decomposition (i.e. adding artist and genre embeddings), there is virtually no difference between `v0' and `v1' semantic IDs (Fig. \ref{fig:results_merged}c).
This may be because artist and genre embeddings perform a similar function to the semantic IDs (i.e. they are also shared between `similar' songs). However, a model trained with only genre and artist embeddings has a much lower accuracy (by $\sim$10\%), suggesting that 
the semantic IDs (whether random or trained) still contribute to the model accuracy.
This may be caused by improved model regularization, similar to findings by \cite{petrov2024, lan2020}.
% however the song residual embedding is still doing *something* because the accuracy of artist+genre only (without song) is lower.
\subsubsection{Increasing model complexity} \label{sec:varying}

% \begin{figure*}[t]

\begin{figure*}[t]
 \includegraphics[width=0.85\textwidth]{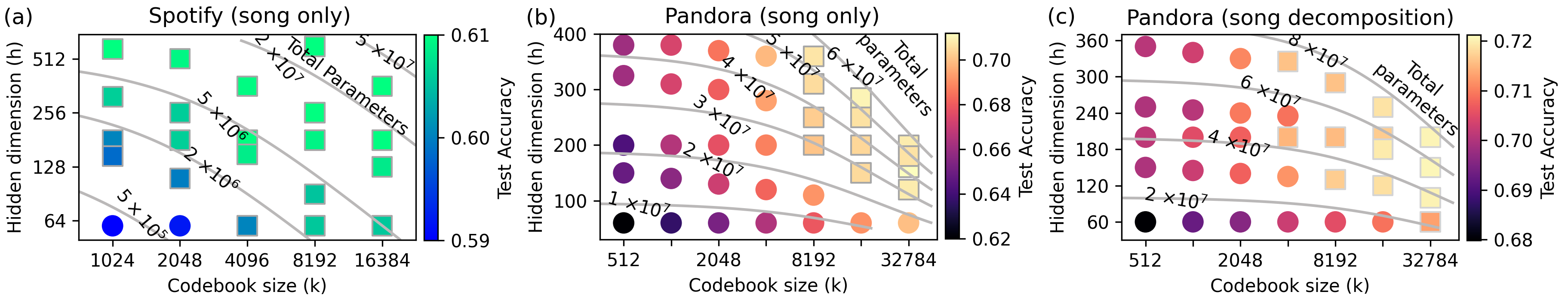}
\caption{
Increasing the number of trainable parameters, either by increasing codebook size $k$ or hidden dimension $h$ improves the test accuracy. For the Spotify dataset (a), the test accuracy generally increases with more trainable parameters, whereas the Pandora dataset (b, c) shows a preference for increasing $k$ over $h$. 
Contours show combinations of $k$ and $h$ with the same total number of trainable parameters. 
The baselines (from Fig. \ref{fig:results_merged}) are trained with $h=60$. Baseline-beating runs are shown as squares.}
  \label{fig:tradeoff}
  \Description{A figure showing the trade-off between increasing codebook size (k) and hidden dimensions (h). In general, for the Spotify dataset (panel a) there is no tradeoff, i.e. the accuracy generally increases with increasing number of parameters, whether these come from k or h. For Pandora (panels b and c) there is a clear preference for higher codebook size. In both cases, certain configurations are able to beat the baseline accuracy.}
\end{figure*}

Because semantic IDs are shared between tracks, this means that fewer parameters are necessary at training and inference. In turn, this allows for more complex models, including at training time, by reassigning those parameters.
There is a trade-off to be made between increasing codebook size $k$ vs. increasing the number of hidden dimensions $h$, as both increase the trainable parameters.
For the Spotify dataset  (Fig. \ref{fig:tradeoff}a), the accuracy in general increases with more trainable parameters, whether these are assigned to increasing $h$ or $k$. 
For the Pandora dataset, both without (Fig. \ref{fig:tradeoff}b) and with (Fig. \ref{fig:tradeoff}c) song decomposition, there is a clear preference for higher $k$ instead of $h$ for a given number of trainable parameters. 
Further work could look into reasons for this difference; for example, we note that the feedback types are very different (implicit play/skip for Spotify vs explicit thumb up/down for Pandora). 
It is also worth testing whether the hierarchical structure of the semantic IDs may allow for smaller codebooks for earlier semantic layers, enabling further parameter reduction.
% Model complexity and accuracy can be increased by increasing either the semantic codebook size $k$ or the model's hidden dimension $h$. 

% \end{figure}

\subsubsection{User input length}
\begin{figure}[t]
 \includegraphics[width=0.475\textwidth]{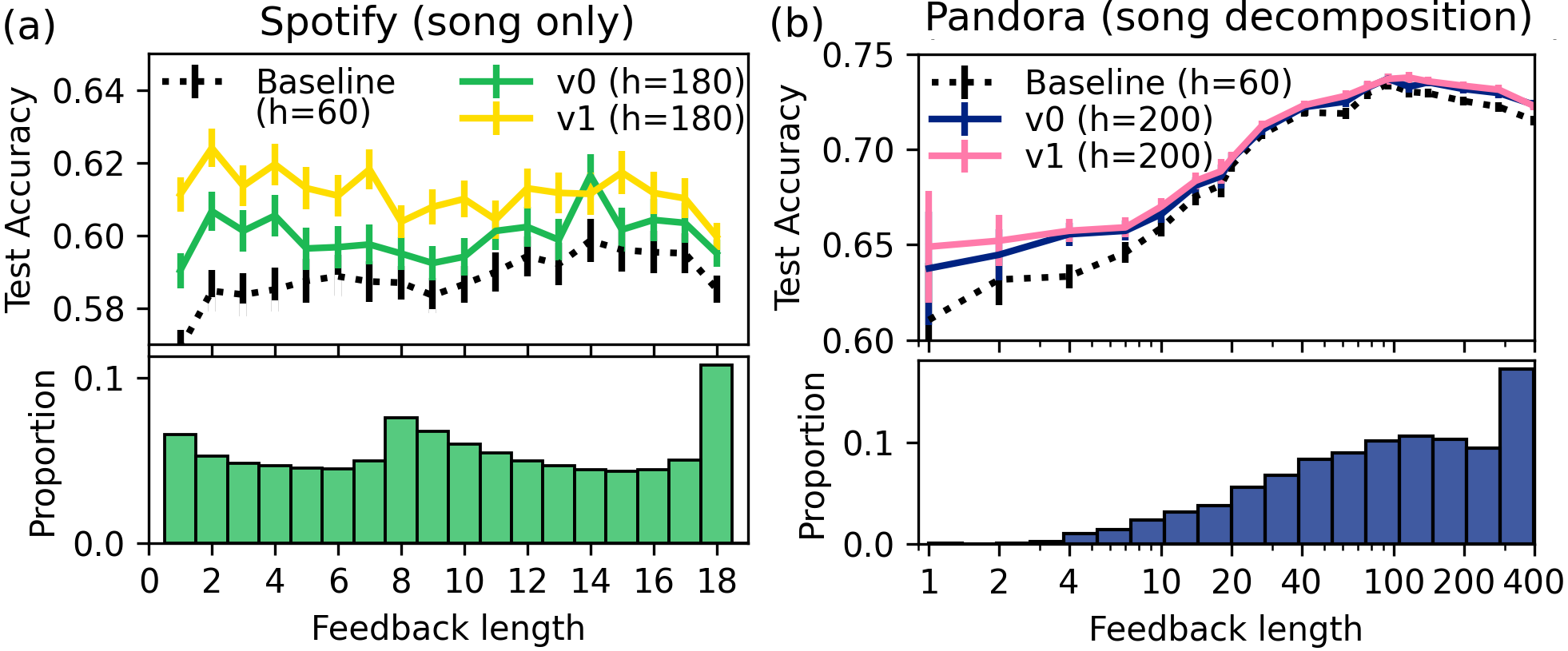}
\caption{The lift in accuracy is highest for low-feedback users for the Spotify (a; $k=4096$) and Pandora (b; $k=16384$) cases. The lift of the (trained) `v1' semantic IDs over the random `v0' ones (for fixed $k$ and $h$) is much larger for the song-only case (a) than when song decomposition is used (b).  Error bars represent 90\% confidence intervals.}
  \label{fig:lengths}
\Description{A figure the difference in accuracy for the baseline, semantic v0 and semantic v1 models for Spotify (panel A) and Pandora (panel B). Both datasets show higher lift for low-activity users (i.e. users with shorter input sequences), with no negative lift for any user segment. The Spotify data has much shorter sequence length (from Table 1) and is curated to have a roughly uniform distribution, whereas the Pandora data is skewed towards lower-feedback users.}
\end{figure}

The Pandora dataset shows greatest lift for low-feedback users; there is also some lift for very high-feedback users (Fig. \ref{fig:lengths}b). 
The Spotify dataset shows lift for all session lengths, with somewhat more lift for shorter session lengths (Fig. \ref{fig:lengths}a).
Just as in Fig. \ref{fig:results_merged}, the `v1' lift in accuracy over `v0' is much larger in the song-only case.  
% For both datasets the lift is highest for low-feedback users/sessions; for the remaining users, Pandora shows some lift for very high-feedback users whereas Spotify shows a fairly consistent lift for most other feedback lengths . 

% \vspace{-0.5em}
\subsection{Online test}
\enlargethispage*{12pt}

\begin{table}[h]
\caption{\label{table:abtest}Online A/B test results ($p$-values shown in brackets).}
\begin{tabular}{@{}l|cc@{}}
\toprule
\multicolumn{1}{c|}{\textbf{Metric}} & \multicolumn{1}{c}{\textbf{Change}} & \multicolumn{1}{c}{\textbf{$p$-value}} \\ \midrule
Listening hours      & -0.08\%  & 0.53 \\
Song completion rate & -0.04\% & 0.22 \\
New releases (<120 days) played & +0.81\% & $\ll10^{-4}$ \\
Distinct songs per seed    & +1.82\% & $\ll10^{-4}$  \\
Distinct artists per seed & +0.51\% & $\ll10^{-4}$  \\ 
Track repetition    & -1.26\% & $\ll10^{-4}$  \\ \bottomrule
\end{tabular}
\end{table}
% \begin{table}[bth]
% \centering
% \caption{\label{table:abtest}Online A/B test results ($p$-values shown in brackets).
% } 
% \begin{tabular}{c|c|c|c}
% \toprule
% \textbf{Metric} & Listening hours &  Active users  & Track repetition   \\ \midrule 
%   \textbf{Change} & $-0.01\%$ (0.93)   & $+0.1\%$ (0.07)       &$-1.2\%$  ($10^{-11}$)   \\

% %Inequality (Gini) of track plays &  $-$0\%    \\
% % Distinct tracks recommended per station & $+$0\% \\
% % Distinct artists recommended per station & $+$0\% \\
% \bottomrule 
% \end{tabular}
% \end{table}

An online test with 10 M Pandora listeners was conducted for 30 days, split into equal control (baseline) and test (semantic) segments (Table \ref{table:abtest}).  The semantic model used had $k = 16384$ and $h = 120$, giving a $\sim$50\% reduction of trainable parameters. 
This allowed for a smaller memory footprint for training, reducing training costs by $\sim$20\%. Although key business metrics were neutral overall, halving the number of trainable parameters with the semantic model and increasing the recommendation diversity are positive outcomes.

\begin{figure}[h]
 \includegraphics[width=0.485\textwidth]{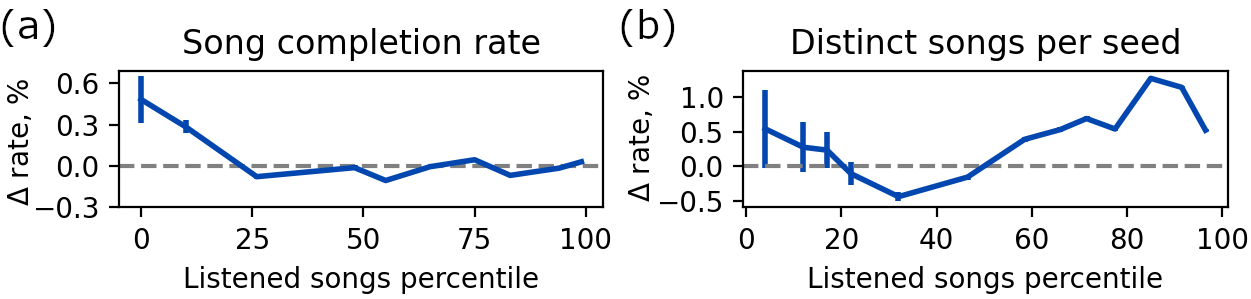}
\caption{The change in song completion rate (a) and the change in the average number of distinct songs playing on each seed (b) for different user segments during the online test period. The x-axis shows users ranked by how many songs they listened to during the test period, with 100 being the highest. Error bars represent 99\% confidence intervals.}
  \label{fig:completion}
\Description{A figure showing the real online test impact on recommendation diversity for Pandora users, cut by listener activity (i.e. how many songs listened to during the test period of 1 month). For both song completion rate (panel a) and distinct songs per seed (panel b), there is a high lift for low-activity users. For song completion, all other user segments show negligible change. For distinct songs per seed, there is also an increase for high-activity users.}
\end{figure}

The (lack of) change in metrics is not uniform across all users. For example, there are significant increases in song completion rate for low-activity users (Fig. \ref{fig:completion}a), though such users by definition do not contribute much listening time overall and hence the overall completion rate change is neutral. 
Nevertheless, this lift for lower-activity users is consistent with the offline estimation from Fig. \ref{fig:lengths} and may give longer-term benefits, if low-activity users are persuaded to become high-activity users due to a better listening experience.

The change in the number of distinct tracks played per seed is also greater for listeners who have very low or very high listening activity (Fig. \ref{fig:completion}b). This may account for the significant reduction in track repetition reported in the online test (Table \ref{table:abtest}). Again, this indicates that the impact of the online test is non-uniform on different user segments.
As a bonus, this increased track diversity did not significantly reduce overall listening hours. 
Future tests may further increase the model complexity and explore the impact of covering cold-start songs using interpolated semantic IDs.

% \begin{table}[bth]
% \centering
% \caption{\label{table:abtest}Online A/B test results.
% } 
% \begin{tabular}{c|cc}
% \toprule
%                  &  Change &$p$-value \\ \midrule 
% Listening hours &   $-0.01\%$  &  0.93  \\
% Active users    & $+0.1\%$       & 0.07    \\
% % Thumb-up rate    & $-0.3\%$      & 0.05    \\
% Track repetition    & $-1.2\%$    & $\ll$ 0.01     \\
% %Inequality (Gini) of track plays &  $-$0\%    \\
% % Distinct tracks recommended per station & $+$0\% \\
% % Distinct artists recommended per station & $+$0\% \\
% \bottomrule 
% \end{tabular}
% \end{table}

% \section{Discussion}

% \subsection{Cold start}
% Out-of-vocabulary songs do not have a song embedding, but would have additional content features that enable coverage (e.g. mapped to an existing set of semantic IDs, or they can be partially covered via artist and genre embeddings). To simulate this effect for (popular) new releases, songs released in the final month of the Pandora dataset were removed from training, but kept in the test set. 
% The test accuracy for these songs was 0.62 when using semantic IDs vs. 0.60 for using an artist + genre fallback vs. 0.5 for random guessing vs. 0.77 if these songs had a trained embedding.
% This suggests that interpolating these new songs onto semantic IDs can provide reasonable temporary coverage, before these songs receive some feedback and can be covered by the model.
% Further work can be done, including an online test, to verify this.

\section{Conclusion}

\enlargethispage*{16pt}
Semantic IDs are a viable way to share content features and reduce model parameters without reducing accuracy. 
This allows for more complex models that may boost model accuracy and diversity and can reduce training/inference compute requirements/costs.
The accuracy lift is highest for low-feedback users, and in general the effects of the semantic IDs are not uniform across different user segments.
Further work could explore the impact of interpolating semantic IDs for new (cold-start) songs.

\section{Speaker Bio}
Jeffrey Mei is a Staff Scientist at Sirius XM / Pandora, specializing in large-scale sequential recommender systems for online radio next-song recommendation. Prior to this, he was a Senior Scientist at Wayfair working on next-item furniture recommendations. He is particularly interested in quantifying qualitative attributes like style and how they may inform cross-category recommendation.

% \begin{acks}
% Blah
% \end{acks}


\bibliographystyle{ACM-Reference-Format}
\bibliography{stamper-bib}

\end{document}